\documentclass[aip,rsi,reprint]{revtex4-1}
\usepackage{graphicx,hyperref,pdfpages}
\usepackage{filecontents}
\usepackage{epstopdf}
\draft 

\begin{document}
\title{Design and performance of an ultra-high vacuum scanning tunneling microscope operating at dilution refrigerator temperatures and high magnetic fields} 

\author{S. Misra}
\author{B. B. Zhou}
\author{I. K. Drozdov}
\author{J. Seo}
\author{L. Urban}
\author{A. Gyenis}
\affiliation{Joseph Henry Laboratories and Department of Physics, Princeton University, Princeton, New Jersey 08544}
\author{S. C. J. Kingsley}
\author{H. Jones}
\affiliation{Oxford Instruments Omicron Nanoscience, Tubney Woods, Abingdon, Oxfordshire, OX13 5QX, United Kingdom}
\author{A. Yazdani}
\email{yazdani@princeton.edu}
\affiliation{Joseph Henry Laboratories and Department of Physics, Princeton University, Princeton, New Jersey 08544}

\date{\today}

\begin{abstract}
We describe the construction and performance of a scanning tunneling microscope (STM) capable of taking maps of the tunneling density of states with sub-atomic spatial resolution at dilution refrigerator temperatures and high (14 T) magnetic fields. The fully ultra-high vacuum system features visual access to a two-sample microscope stage at the end of a bottom-loading dilution refrigerator, which facilitates the transfer of {\it in situ} prepared tips and samples. The two-sample stage enables location of the best area of the sample under study and extends the experiment lifetime. The successful thermal anchoring of the microscope, described in detail, is confirmed through a base temperature reading of 20 mK, along with a measured electron temperature of 250 mK. Atomically-resolved images, along with complementary vibration measurements, are presented to confirm the effectiveness of the vibration isolation scheme in this instrument. Finally, we demonstrate that the microscope is capable of the same level of performance as typical machines with more modest refrigeration by measuring spectroscopic maps at base temperature both at zero field and in an applied magnetic field.
\end{abstract}

\pacs{07.79.Cz,07.79.Fc}
\maketitle 

Scanning tunneling microscopy (STM), since its development almost 30 years ago, has become a powerful technique in condensed matter physics, providing not only structural information about surfaces, but also spectroscopic measurements of the electronic density of states at the atomic length scale. However, most instruments operate at temperatures above 1 K, limiting access to exotic electronic phases and quantum effects expected at lower temperatures, which are studied as a matter of routine by other techniques. Generally, very little spectroscopic information about the electronic density of states is known at dilution refrigerator temperatures, usually being limited to what can be learned using either point contact spectroscopy or planar tunnel junctions. Moreover, STM can make such measurements on the atomic length scale, allowing it to probe systems, such as single spins and atomic chains, which are not directly accessible any other way. 

While the integration of STM with a dilution refrigerator can be conceptually reduced to simply attaching the microscope to the end of a mixing chamber in lieu of some other cryogenic refrigerator, the technical requirements for sub-Angstrom positioning of an STM tip above an atomically clean surface are often at odds with those for cooling a sample to milli-Kelvin temperatures. For example, when attaching the microscope to the refrigerator, the former would favor the use of a soft mechanical joint using springs, which would isolate vibrations, while the latter would favor the use of a rigid mechanical joint with a metal rod, which would provide a strong thermal contact. Nevertheless, a number of STM instruments have been developed that cool the sample using a dilution refrigerator \cite{PannetierRSI,Barker,FukuyamaRSI,GollRSI,StroscioRSI,WahlRSI,SuderowRSI,KernRSI}. However, among these, few feature ultra-high vacuum (UHV) environments, which would facilitate the {\it in situ} preparation of tips and samples, a crucial step in preparing many samples and functionalizing STM tips \cite{KernRSI,FukuyamaRSI,StroscioRSI}. Moreover, few have the level of stability and performance required to measure spectroscopic maps of the electronic density of states with atomic spatial resolution, crucial to obtaining detailed information about the electronic state of the compound under study \cite{Barker, SuderowRSI, FukuyamaRSI, StroscioRSI}. Here, we describe the construction and performance of a home-built STM designed specifically to extend the level of functionality and stability common in higher temperature systems to dilution refrigerator temperatures.

\section{Ultra-high vacuum assembly}

\begin{figure*}
\includegraphics{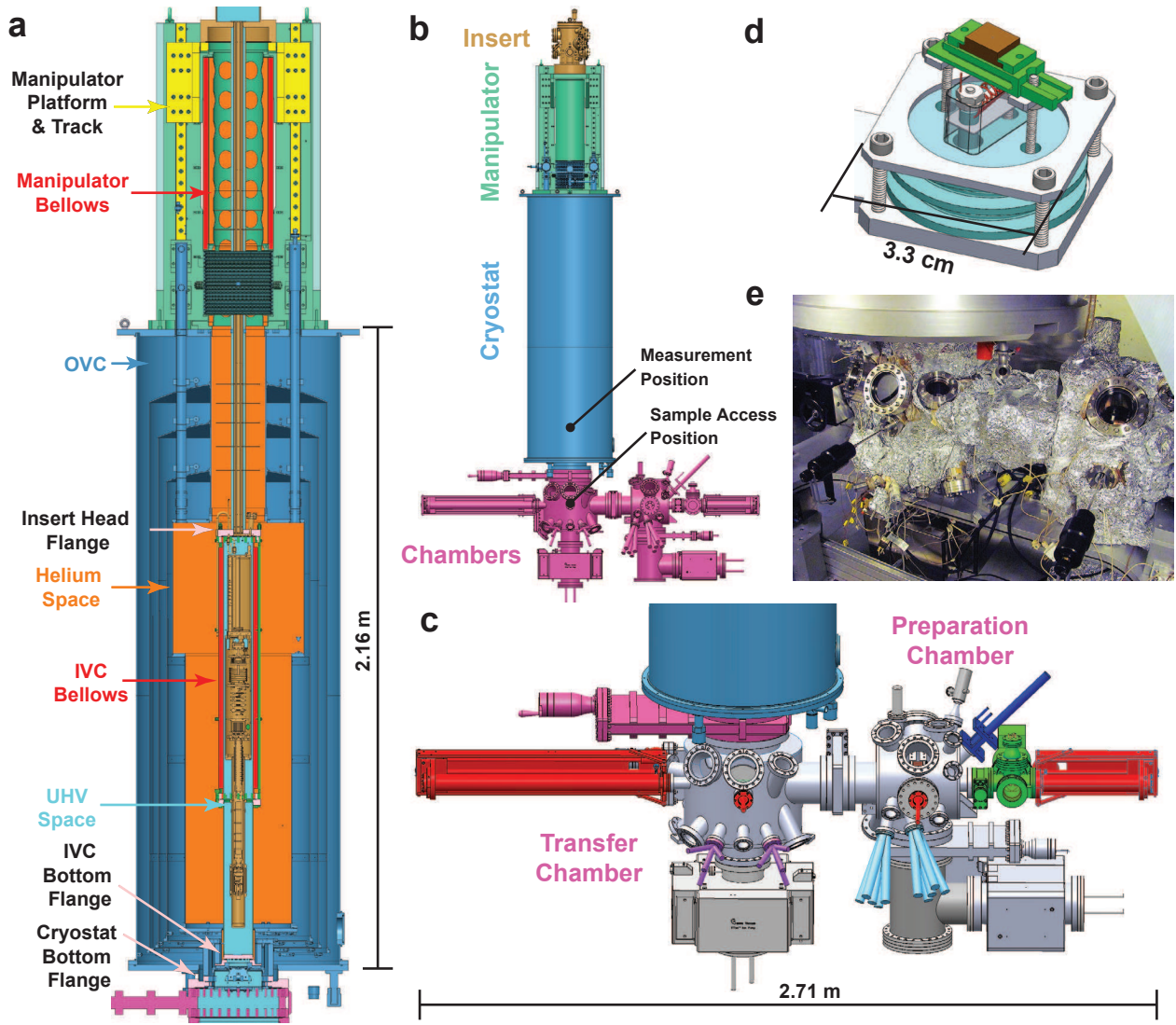}
\caption{(a) This 3D CAD drawing shows a zoomed-in cross-section of the insert, manipulator and cryostat, whose connection to the UHV chambers is shown in (b). When the manipulator platform moves down its track (yellow), the manipulator and IVC bellows (red) contract, and the insert (brown) moves out the end of a UHV neck at the bottom of the cryostat into the chambers below, translating the attached microscope between the measurement and sample access positions labeled in (b). The relevant Conflat flanges which interface the insert to the IVC (IVC head flange), the IVC to the cryostat (IVC bottom flange), and the cryostat to the vacuum chambers (cryostat bottom flange) are shown in pink. (c) The UHV chambers beneath the cryostat include a transfer chamber, a preparation chamber, and a load lock (green). The wobble sticks and manipulators used to transfer samples are shown in red. The focal point of the transfer chamber can accommodate a number of evaporation sources (two are shown in purple). The left-hand focal point of the preparation chamber has a resistive heater stage for samples or tips (not shown), and can have up to four evaporators pointed at it (two are shown in cyan). The right-hand focal point of the preparation chamber has an e-beam heater stage with an Ar ion sputter gun pointed at it (dark blue). (d) This drawing shows a close-up of the e-beam heater stage, with the sample holder shown in green. Alumina pieces are shown in light blue, and the filament in red. (e) Photograph of the transfer and preparation chambers attached to the cryostat.
}
\label{ga}
\end{figure*}

The successful integration of a dilution refrigerator into an ultra-high vacuum environment has the unmeasurable benefit for scanning tunneling microscopy that the full range of samples available to the technique could be studied, and standard techniques for the {\it in situ} preparation of tips and samples could be used without alteration. Although standard dilution refrigerators contain materials, such as nylon, brass and soft solder, which are anathema to ultra high vacuum, substitution by UHV-compatible materials (PEEK, OFHC copper, and high temperature solder) and the adoption of proper cleaning methods has been successfully implemented in a number of systems \cite{Goldman, Shvarts, Kingsley}. The remaining difficulty lies in devising a scheme to transfer tips and samples between various UHV stages and the microscope.

Toward this end, we mount a specially designed bottom-loading dilution refrigerator insert onto a z-manipulator which can translate the insert and attached microscope between the measurement and sample access positions within contiguous UHV space (Fig.~\ref{ga}b). This UHV space of the instrument extends up from the chambers via an inner vacuum can (IVC) with a flexible bellow up to a head flange on the insert (Fig. ~\ref{ga}a). With the exception of the wiring interface, which is through a chamber at the top of the insert and connects to the head flange through a series of tubes, the head flange is the terminus of the UHV space. The top of the rigid insert is secured to a heavy duty (non-UHV) z-manipulator. This manipulator lowers the entire insert up to 65 cm, collapsing the IVC bellows. This, in turn, moves the microscope, which normally sits at the center of a 14 T magnet (103 mm bore diameter) when the manipulator is up, through the bottom neck of the cryostat, and produces it at the center of the UHV chamber below. After opening a rotary door on the radiation shield of the refrigerator using a multi-motion wobble stick, we have direct visual access to the microscope itself. 

The UHV utilities in the three chambers attached to the cryostat have been specifically designed to enable the implementation of the full suite of recipes for {\it in situ} preparation of spin-polarized STM tips and samples\cite{Roland}. New tips and samples are introduced into the UHV chambers through a standard load lock attached to the preparation chamber (Fig. \ref{ga}c).  The preparation chamber contains two points which each lie at the focus of multiple ports of the vacuum chamber, one of which has an e-beam heater (Figure 1d) and sputter gun to clean tips and metal samples, and the other of which has a resistive heating stage and evaporators which can be used to grow thin metal films on them. To allow for evaporation onto a cold sample, the transfer position of the dilution refrigerator insert sits at the focal point of three ports of the transfer chamber, to which standard evaporators can also be attached. Finally, cleavable samples can be both cleaved and stored in the preparation chamber.

The operating base pressure of the system is $\sim 10^{-10}$ torr. The UHV chambers can achieve this level of vacuum simply by baking to 130 C for two days. The insert has a pair of heaters located near the microscope, but can only be baked to 60 C. Despite this limitation, after cooling to liquid helium temperatures, the insert does not change the level of vacuum in the transfer chamber, even when swapping samples or tips. As shown in the last section, this level of vacuum is more than sufficient to leave prepared surfaces clean for examination with the microscope.

\section{Vibration isolation}

\begin{figure*}
\includegraphics{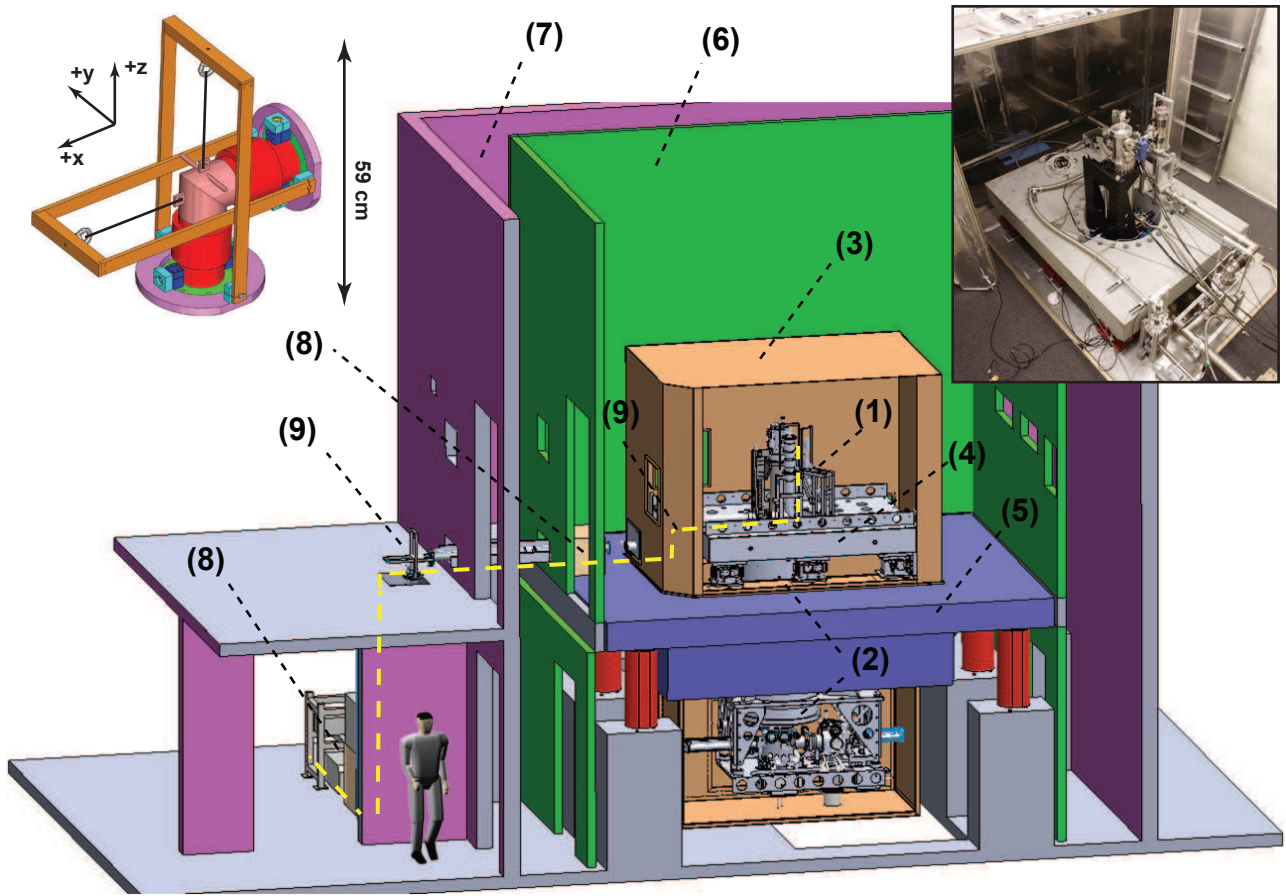}
\caption{The instrument (1) is mounted to a custom-made aluminum support structure (2), both of which are surrounded by a radio frequency (RF) isolation enclosure (3). In order to facilitate the mating of the insert to the cryostat, which requires moving the chambers out of the way, the chambers are mounted to an aluminum table, which is attached to the aluminum frame by way of special kinematic mounts. This entire support structure is suspended freely by securely bolting to a 4 ton granite table (4) above. This table floats on a set of 6 passive isolators, and serves as one isolated stage. This, in turn, rests on a 30 ton concrete plinth (5), which itself sits on a set of another 6 heavy passive isolators, and defines a second isolated stage. An acoustic enclosure (6) surrounds the entire structure. The plinth, the acoustic enclosure, and an external grout-filled concrete wall (7) connect only through the floor of the basement lab. A concrete block (8) and gimbal (9) is used to isolate vibrations which might be transmitted by the pumping lines between the floor and the plinth, and again between the plinth and the granite table (pumping lines delineated by yellow line). (left inset) This shows a 3D CAD drawing of our double gimbal, whose purpose is to decouple the motion of the two flanges (green), which are each attached to rigid pipes. This is accomplished by attenuating the motion of either of the flanges with respect to the central elbow (pink) through the use of two sets of edge welded bellows (red). The position of this elbow is determined by vacuum forces balanced by the tension on wire ropes (black) connected to the arms (brown). If the pipe attached to the bottom flange of the assembly were to move in +z (+x), then the horizontal (vertical) arm would twist upwards (right) on an axial bearing (light blue). Similarly, if it were to move in +y, the base plate (purple) would twist to accommodate the motion. (right inset) Aerial photograph of the instrument with detachable RF room top moved aside.
}
\label{isolation}
\end{figure*}

\begin{figure}
\includegraphics{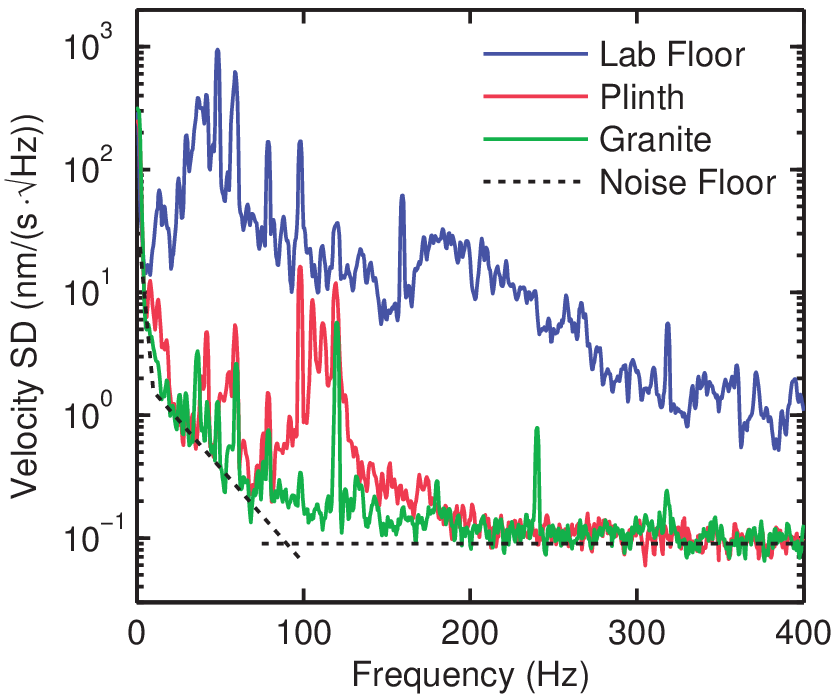}
\caption{This plot shows the velocity spectral density, as measured by a Wilcoxen 731A, present on the lab floor (blue), on top of the plinth (red), and on top of the granite slab (green). The combined baseline sensitivity of the accelerometer and spectrum analyzer is shown as the noise floor (black). The data were taken while running the dilution refrigerator.
}
\label{vibration}
\end{figure}

The quality of the data taken by any STM is largely determined by its ability to limit the level of vibrations in the tip-sample junction, to which the tunnel current is exponentially sensitive. For our microscope, the need for strong thermal coupling to the dilution refrigerator precludes the use of springs at the microscope itself, a common technique which is remarkably effective. Instead, as we outline in Figure ~\ref{isolation}, we have isolated the entire instrument shown in Figure ~\ref{ga}b from external acoustic and floor-borne sources of noise. Vibrations present in the floor of the laboratory are first attenuated down to very low frequencies ($\mathrm{\sim 1 Hz}$) by a set of six passive pneumatic legs which float a 30 ton concrete plinth, as shown in Figure ~\ref{vibration}. A passively isolated 4 ton granite slab sits on top of this concrete plinth, providing an additional layer of isolation from floor-borne noise. To realize this low level of vibrations in the microscope itself, the instrument (cryostat and chambers) is secured to a dissipative heavy duty aluminum frame, whose only rigid attachment is to this vibration-isolated granite slab above. A similar two-tiered scheme was realized to attenuate acoustic vibrations. Two layers of acoustic shielding are realized by surrounding the plinth first by an acoustic enclosure, and then surrounding that enclosure by a second room built from grout-filled concrete blocks. 

This scheme to isolate vibrations can be rendered useless unless proper care is taken in handling the large pumps and pumping lines required to run a dilution refrigerator. The pumps generate a lot of noise, and the pumping lines not only transmit these vibrations, but their stiffness can mechanically short the pneumatic isolation stages together. To attenuate high frequency vibrations, the four gas lines, which includes the still and 1K pot pumping lines, the condenser and the cryostat exhaust, are cast in a 0.5 ton concrete block located on the lab floor (Fig. ~\ref{isolation}). To attenuate low frequency vibrations and prevent mechanical shorting of subsequent pneumatic stages, the 1K pot pumping line, the condenser and the dewar exhaust have long looped sections of formed bellows between the lab floor and the plinth. To achieve the same effect on the much larger diameter (and thus much stiffer) still pumping line, a double gimbal based on the design of Ref. \onlinecite{Gimbal} (left inset of Fig. ~\ref{isolation}, manufactured by Energy Beams, Inc.) is used to bridge the gap between the lab floor and the plinth. This combination of concrete block and gimbal is repeated again when going from the plinth to the granite table. As shown in Fig. ~\ref{vibration}, the end result of our isolation scheme establishes sub-nm/(s$\cdot\sqrt{\mathrm{Hz}}$) vibration levels approaching the noise floor of our accelerometer for a wide range of frequencies even with all pumps attached and running. Having created a suitably low vibration environment, the STM, which we describe in the next section, must be made as rigid as possible to realize the level of performance required to take low noise measurements, which we describe in the last section.

\section{The dilution refrigerator and microscope head}

\begin{figure*}
\includegraphics{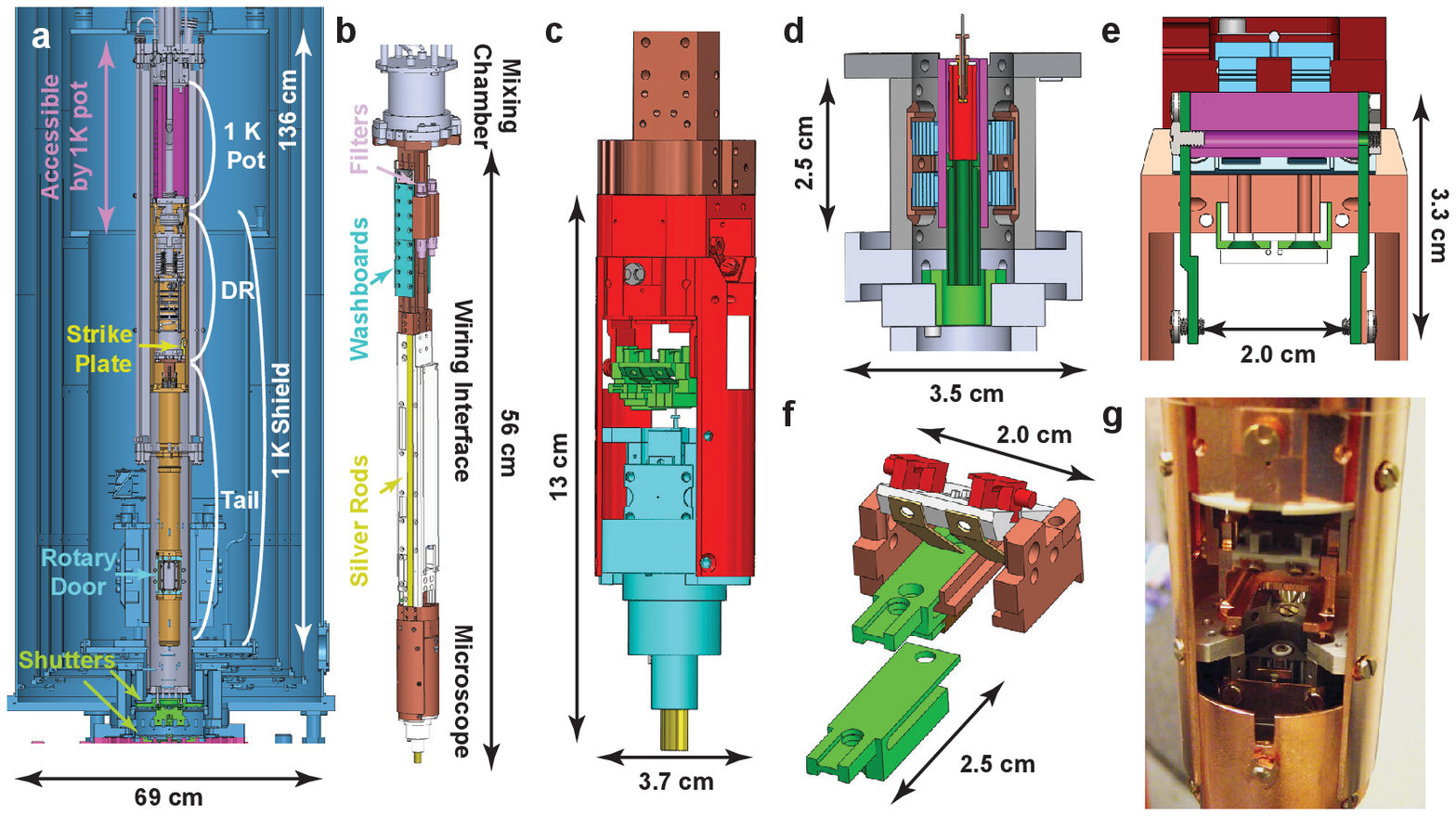}
\caption{
(a) This 3D CAD cross-section of the general assembly highlights the non-standard features of the insert. The oval cam and strike plate used to pre-cool the microscope are shown in yellow. The approximate volume of the main bath accessible to the 1K pot pickup line is shown in purple. The rotary shutter on the 1K radiation shield is shown in cyan. The two sets of radiation baffles on the cryostat, which sit at roughly 77 K and 4K, are shown in green. The location of the 1K pot (purple), the 1K radiation shield, the tailpiece (tail), and dilution refrigerator (DR) are also labeled. (b) This is a 3D CAD drawing of STM tailpiece, highlighting the washboards (cyan), the tip and sample RF filters (purple), and the silver rods (yellow). (c) This shows a 3D CAD drawing of the microscope, with the z-motor shown in blue, the x-motor shown in red, and the sample cubby shown in green. A radiation shield which shields the tip and the sample is hidden from view. (d) A cross-section of the z-motor is shown here, with the prism in purple, the scan piezo in red, piezo stacks in blue, the bottom plug in dark green, and the capacitance sensor in light green. (e) A cross-section of the x-motor, with the prism in purple, the piezo stacks in blue, the arms which hold the sample cubby in dark green and the capacitance sensor plates in light green. (f) This is a 3D CAD model of the sample cubby, showing the samples (green) and the PEEK lid (white). This lid, when pushed down, compresses two BeCu springs on the backs of the samples. The two dovetail pieces (red) ride on tracks, and can lock the lid into place. (g) Photograph of the microscope with the milli-Kelvin radiation shield pulled down to reveal access to the empty tip receptacle and sample cubby (the CAD in (f) shows a newer version of the cubby than the photograph does).
}
\label{insert}
\end{figure*}

The last challenge is the conceptually simple step of attaching an STM head to a dilution refrigerator, with the goal of attaining the lowest possible temperature while retaining the maximum amount of functionality in the STM head. Thermally, the insert on our system is a fairly standard design, but with four notable exceptions (Figure ~\ref{insert}a) to accommodate the UHV compatibility of the system. First, given that the IVC is a UHV space, exchange gas cannot be used to cool the insert from high temperatures down to 4 K, either for the initial cooldown of the system, or when transferring samples. Instead, we have an oval-shaped mechanical heat switch, operated using a rotary feedthrough, which allows the mixing chamber to be thermally shorted to the 1K shield via mechanical contact. This allows us to cool the system from room temperature to 2 K in around 40 hours during the initial cooldown, and from around 40 K to 2 K in around 6 hours when transferring samples. Second, because the insert needs to be moved up and down and the 1K pot has a fixed length pickup tube, about half the helium in the main bath is accessible to the pot, resulting in shortened time between refills of the main bath. That the cryostat has two satellite necks (to accommodate the demountable magnet current lead and cryogenic services) exacerbates this by increasing the helium consumption. Still, the helium consumption rate with the dilution refrigerator running is around 18 liters a day, resulting in a hold time for the main bath of 4 days. Finally, in order to have access to the microscope inside UHV, the radiation shields in the cryostat and the 1K shield need to be able to be opened inside the vacuum space, which could compromise their performance. These appear not to introduce any unexpected radiative heat load, as the dilution refrigerator was measured to achieve a base temperature of 8 mK (measured using a cobalt-60 nuclear orientation thermometer) and had a cooling power of 400 $\mu W$ at 100 mK with only the thermometry installed. 

To cool the microscope to milli-Kelvin temperatures, we borrow standard techniques used for sample-in-vacuum dilution refrigerator instruments (Figure ~\ref{insert}b). The primary heat load added when installing an instrument on the mixing chamber comes from the wiring. Both the shielded and unshielded lines extending down from room temperature are thermally anchored first at 4K, then at the 1K pot using 3 cm long washboards, and then connected to either flexible stainless/ NbTi coaxial cable (custom from New England Wire Technology) or NbTi wire. Because NbTi superconducts at 1K pot temperatures, these effectively act as a thermal break between the 1K pot and the mixing chamber. Between the mixing chamber and the microscope, however, we would like to maximize the thermal conductivity, and accordingly we use silver-coated copper coaxial lines and wires that are anchored using 10 cm long washboards screwed tightly into a copper stub that is attached to the mixing chamber. In addition, the tip and bias lines are fed through a lumped element RF filter (VLFX-1050+ from Mini Circuits) at the mixing chamber to filter out unwanted high frequency noise in the tunnel junction. To efficiently cool the body of the microscope itself, which is 42 cm away from the mixing chamber, we link the two with silver rods (3N5 purity, $\mathrm{50 mm^2}$ in cross-section, from ESPI Metals, Inc.) which have been annealed \cite{Ian} to enhance their thermal conductivity. These fit in a PEEK frame which houses electrical connectors and serves as a secure mechanical attachment point for the microscope. With the microscope installed, the base temperature of both the mixing chamber and the microscope was measured to be 20 mK using a ruthenium oxide thermometer anchored at the microscope. The mixing chamber now retains 260 $\mu W$ of cooling power at 100 mK, leaving sufficient flexibility for adding more lines to the system when more elaborate experiments need to be done.

The microscope (Fig. ~\ref{insert}c) contains three functional blocks: a z-axis piezo motor, an x-axis piezo motor, and a two-sample cubby. Both the motors are Pan walkers \cite{Pan}, in which a triangular sapphire prism (custom, from Swiss Jewel Company) can be moved along a single axis of motion using three pairs of piezoelectric shear stacks (Model P121.01H from Physik Instrumente L. P.) in a slip-stick motion. The bodies of both the motors are made of OFHC copper, coin silver, and PEEK pieces held together using non-magnetic silicon bronze screws (custom, from Swiss Screw Products, Inc.). The z-motor (Fig. ~\ref{insert}d), which provides a total of 4 mm of vertical motion, is used to approach and withdraw a scan piezo (3 \AA/V sensitivity in z, 9.5 \AA/V sensitivity in x/y, EBL \#4 material from EBL Products, Inc.) which is glued to the inside of a cylindrical cavity in the prism. This motor reliably produces 80 nm-sized approach steps at low temperatures with a drive voltage of 375V supplied by a Nanonis PMD4 piezo motor driver. Its absolute position can be tracked using a cylindrical capacitor formed by a metal end-plug in the prism and a corresponding piece which is part of the (static) microscope body. The x-motor provides a total of 7 mm of horizontal motion in 280 nm-sized steps to the attached sample cubby (Fig. ~\ref{insert}e). A pair of capacitors, each formed by the arms that carry the sample cubby and two internal (static) plates on the microscope body, allow us to track the absolute horizontal position of the sample. The x-motor has been designed to have such a large offset range specifically to allow us to move the tip between two samples (Fig. ~\ref{insert}f) in the cubby. In addition to being designed to accommodate two samples, the cubby has a PEEK lid, which, when opened using the wobble stick, allows samples to be easily slid into the stage, and, when closed using the wobble stick, compresses two BeCu spring contacts firmly into the samples and locks into place. This mechanism provides a more solid mechanical and thermal joint between the sample holders and the sample cubby than would otherwise be possible. In addition to the two samples, the STM tip can also be swapped {\it in situ} by plugging it with a wobble stick into a BeCu collet.  While the ability to offset the sample over a large range provides considerable conveniences as discussed below, the lowest resonance frequencies of the microscope is likely associated with the pendulum modes of the two arms (molybdenum) attaching the sample cubby to the x-motor (shown in Fig. ~\ref{insert}e).  By exciting the the x-motor piezo stacks with a drive voltage of swept frequency and measuring the response in the current, we determine the strongest resonance of the x-motor and sample cubby assembly to be at $\sim$900 Hz at room temperature, with an additional weaker response at $\sim$700 Hz.

The combination of the two-sample holder cubby, the reliability of the motion of the motors, and the repeatability and precision of the capacitive position sensors provide a critical function when studying samples which cleave poorly. Approaches onto bad areas of the cleavable sample which change the tip do not terminate the experiment, but rather can be simply corrected by field emitting and checking the tip on a simple metal sample in the other slot. After field emission, using the previous position register, we can return to the same spot (macroscopically) on the cleavable sample to continue looking for an acceptable area. Most importantly, even when an acceptable area is found, the microscope can be used to look for an even better area, or for areas with rare surface terminations, with the knowledge that the sensors and motors are reliable enough to allow the microscope to return to the original area should another good area not be found. The ability to systematically search for the best area, or a very rare surface, on a cleavable sample greatly reduces both the number of samples and the time it takes to do an experiment when compared to being limited to examining representative areas, as is the case in most STM instruments.

\section{STM performance}

\begin{figure}
\includegraphics{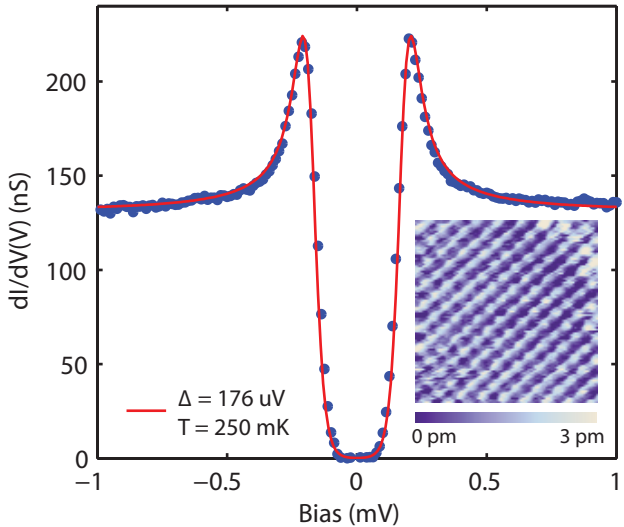}
\caption{
This plot shows the differential tunneling conductance of a superconducting Al(100) sample measured with a normal PtIr tip, using a lockin amplifier with an ac modulation of 5 $\mu V$ at 1.11 kHz, along with a fit to the thermally broadened BCS density of states (red). The spectrum took 8 minutes to acquire at a setpoint of 200 pA at -1.5 mV. (Inset) Unfiltered topographic image at base temperature over 30 $\AA$ at a setpoint current of 1 nA and bias of -1 mV of the same Al sample showing $\mathrm{\sim}$2 pm amplitude atomic modulations.
}
\label{aluminum}
\end{figure}

As we have already described, the instrument provides a similar level of functionality to that present in higher temperature STMs. In this last section, we show that it also provides a comparable level of performance, but at significantly lower temperatures, and in high magnetic fields. All data presented are measured using a Nanonis SPM controller and a Femto LCA current preamplifier with 1 kHz bandwidth and $5 \cdot10^9$ V/A gain. In Figure ~\ref{aluminum}, we show data taken on a Al(100) surface prepared {\it in situ} using ion sputtering and annealing, and measured at base temperature and zero magnetic field. The topograph in the inset shows well-resolved atoms, even on a challenging material where the atomic corrugation is very small ($<$ 5 pm). Moreover, the differential conductance measured on Al provides a direct measure of the electron temperature of samples placed in our microscope. This temperature can exceed the measured lattice temperature (20 mK) due to heating from unthermalized electromagnetic radiation transmitted from room temperature to the sample along the electrical line delivering the sample bias \cite{Vion,Powder}. Fitting our tunneling density of states on Al to the thermally broadened density of states for a Bardeen-Cooper-Schrieffer superconductor $\rho_{BCS}(E) \propto {|E|\over{\sqrt{E^2-\Delta^2}}}$, where $\Delta$ is the size of the gap, we find the size of the gap to be $\Delta =$ 176 $\mu V$, in agreement with that expected for Al(100) tunnel junctions \cite{Algap}, and the electron temperature of our instrument to be 250 mK. Ideas for reducing this electron temperature are discussed in the conclusion.

\begin{figure}
\includegraphics{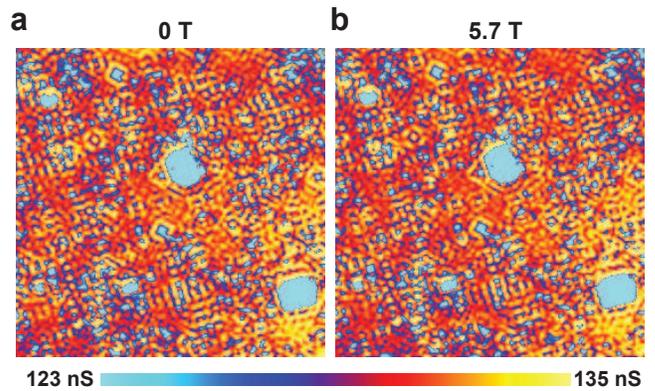}
\caption{This figure shows two spatial maps of the tunneling conductance, each recorded at a bias of +2 mV, over a field of view of 67 nm square on the heavy fermion superconductor $\mathrm{CeCoIn_5}$ \cite{ZhouNPhys}. The maps were taken at base temperature on the same area at (a) zero and (b) 5.7 T applied field using an ac modulation of 66 $\mu V$ at 1.11 kHz. 
}
\label{maps}
\end{figure}

More importantly, the overall level of noise is low enough to enable us to measure one of the most demanding, but most powerful, kinds of data typically taken using STM: spectral maps, in which the STM is used to visualize spatial patterns in the density of states at a fixed energy with sub-Angstrom spatial resolution. In Figure ~\ref{maps}, we show two such maps taken at base temperature on the heavy fermion superconductor $\mathrm{CeCoIn_5}$  \cite{ZhouNPhys}, a system in which superconductivity develops below 2.3 K and can be extinguished upon application of a 5 T field at low temperature. The maps are taken at an energy outside the superconducting gap (measured to be 0.5 mV), and demonstrate that the standing waves created by the interference of scattered quasiparticles, which bear the fingerprint of the underlying band structure, are not altered by the application of a field large enough to suppress superconductivity in this material. These maps also indirectly demonstrate the stability of the microscope, which, over the course of several weeks, was able to take two dozen spectral maps at different energies, both at zero field and in applied field, on the same area of the sample. 

\begin{figure}
\includegraphics{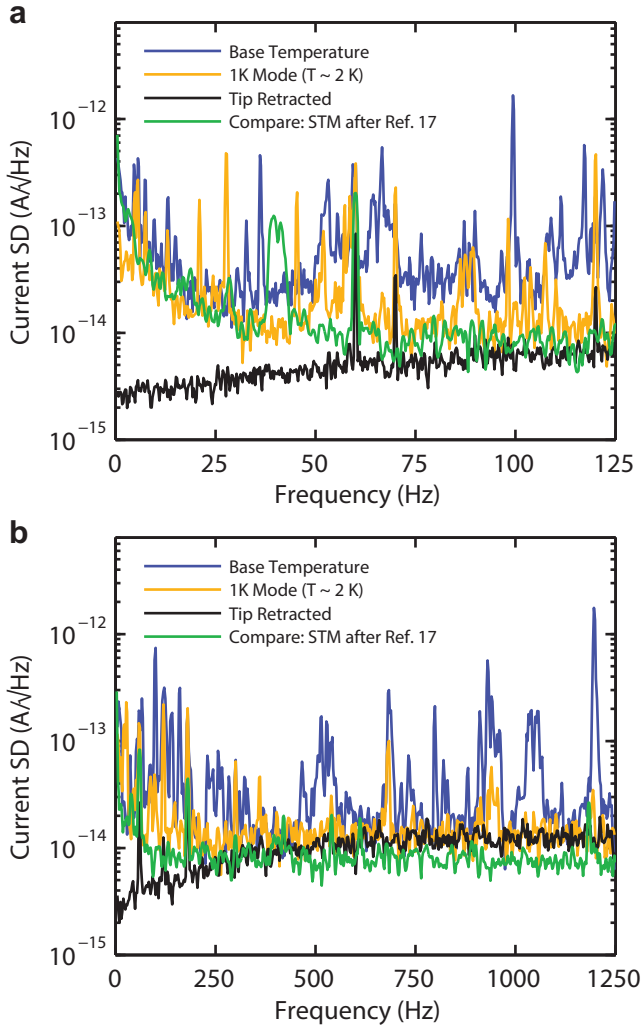}
\caption{
These are plots of the open feedback noise on the tunneling current over two frequency ranges (125 Hz in (a), 1250 Hz in (b)) with the 1K pot running in single shot-mode (orange) and the dilution refrigerator circulating with the 1K pot in single-shot mode (blue). Also shown is the open feedback noise on an STM based on the design of Ref. \onlinecite{VTSTM} (green). The data were taken with a DC tunneling current of 100 pA and a bias of -200 mV on a clean Cu (111) surface. The black curve is the noise on the current with the tip outside of tunneling range, and serves as a measure of the bare amplifier noise due to stray capacitance.
}
\label{currentnoise}
\end{figure}

The data in Figures ~\ref{aluminum} and ~\ref{maps} are qualitative proof of the level of performance of this instrument. A more quantitative figure of merit for any STM, one which can be used to compare the relative levels of performance of different instruments, or different modes of operation, is the integral of the spectral density of the current over low frequencies. The limiting noise in STM measurements is invariably low-frequency noise because it is time-consuming to average out and shows up in both topographic and spectroscopic measurements. In Figure ~\ref{currentnoise}, we compare the spectral noise density of the tunneling current for this instrument and another STM in the lab which uses springs and magnetic damping right at the microscope head itself to isolate vibrations. Notably as shown in Figure ~\ref{currentnoise}a, this instrument has a comparable amount of low frequency noise (in orange; integral over 125 Hz is $\mathrm{0.49\,pA^2}$) to the more conventional design (in green; $\mathrm{0.28\,pA^2}$ over 125 Hz) when the system runs with the 1K pot in single-shot mode and the dilution refrigerator off. The amount of low-frequency noise when also running the dilution refrigerator continuously, while considerably larger (in blue; $\mathrm{2.5\,pA^2}$ over 125 Hz), is directly comparable to published measurements from the UHV dilution refrigerator STM of Ref. \onlinecite{StroscioRSI} ($\mathrm{\sim 3\,pA^2}$ over 100 Hz, using Fig. 19). However, in our case, the system must be run with the 1K pot in single-shot mode, which means it must be refilled every 8 hours, in order to achieve this level of noise. While the spectral noise density contains many seemingly deleterious resonances at higher frequencies, as shown in Figure ~\ref{currentnoise}b, these have a minimal impact on our measurements in practice.  Most occur at frequencies far above the bandwidth of the feedback loop, and hence don't appear in topographs or as a set-point error in spectroscopic measurements.  Moreover, the lockin oscillator frequency can still be set to frequencies where the noise spectrum is no worse than the amplifier background (near $\sim$1100 Hz, for example). Evidence suggests that much of the relevant noise originates in the 1K pot itself. The level of noise with both the dilution refrigerator and 1K pot running single-shot is comparable to that when running only the 1K pot single-shot with the dilution refrigerator off. Conversely, the level of noise when running the 1K pot single-shot and dilution refrigerator continuously is comparable to that when running the 1K pot continuously and leaving the dilution refrigerator off. 

\section{Conclusion}

The recent development of this and other dilution refrigerator STMs opens the door to studying exotic electronic phases and quantum phenomena which only occur at milli-Kelvin temperatures with a spectroscopic tool and at the atomic scale. We have described the construction of an instrument which extends both the functionality and level of performance present in higher temperature STMs down to dilution refrigerator temperatures. As the current measured electron temperature of 250 mK is an order of magnitude larger than the measured lattice temperature of 20 mK, further improvements to thermalizing the tip and sample electron temperature can be made. For example, increasing the cooling power delivered to the sample cubby itself (currently thermally anchored through only a thin strip of silver foil 0.4 mm$^2$ in cross-section to facilitate sample motion) and adding more stages of low temperature RF filters \cite{Vion,Powder} on all of the electrical lines are planned.  Only the tip and bias lines are currently filtered at the mixing chamber; all other lines to the microscope are filtered externally at room temperature, potentially causing heating via cross-radiation at the microscope. Moreover, for ease of access, the system has never been operated with the RF enclosure fully closed, which may be necessary for the lowest electron temperatures. Finally, a reduction of the noise likely created inside the 1K pot would reduce the time it takes to acquire the high resolution data shown here, which is limited to being taken in 8 hour intervals due to the single-shot lifetime of the 1K pot. Ideally, a set of two dozen spectral maps could be taken in a handful of days, as is common in more conventional instruments, instead of a couple of weeks. 

\begin{acknowledgments}
We acknowledge key contributions made by Guido van Loon, Steve Shedd, and Peter Heiland from Integrated Dynamics Engineering in the design and construction of the aluminum support structure, Michael Gaevski and George P. Watson from the Micro Nano Fabrication Laboratory at Princeton University in the annealing of the silver rods, Steven Lowe and William Dix from Department of Physics at Princeton University for machining assistance, Jeff Coles, Paul Busby, Adrian Bircher and Andy Yardy at Oxford Instruments in the design and construction of the cryostat and the dilution refrigerator, and Se-Jong Kahng from Korea University in the design and construction of the STM head. The instrumentation and infrastructure were supported by grants from NSF-DMR1104612, ARO grants W911NF-1-0262 and W911NF-1-0606, the Linda and Eric Schmidt Transformative Fund, and the W. M. Keck Foundation.  
\end{acknowledgments}

\bibliography{DRSTM}
\end{document}